\documentstyle[12pt]{article}
\newcommand{\bq}{\begin{equation}}
\newcommand{\eq}{\end{equation}}
\newcommand{\ba}{\begin{eqnarray}}
\newcommand{\ea}{\end{eqnarray}}

\newcommand{\nn }{ \nonumber  }

\newcommand{\ul}{\underline}
\newcommand{\p}{\partial}

\newcommand{\s}{\sigma}

\newcommand{\B}{\Box\,}
\newcommand{\al}{\alpha}
\newcommand{\be}{\beta}
\newcommand{\de}{\delta}
\newcommand{\ga}{\gamma}
\newcommand{\om}{\omega}
\newcommand{\ep}{\epsilon}
\newcommand{\vs}{\vspace*{0.7cm}}
\newcommand{\ms}{\vspace*{1.2cm}}
\newcommand{\hs}{\hspace*{0.5cm}}
\newcommand{\nind}{\noindent}

\begin{document}
\vspace*{2cm}
{\bf\begin{center}
 NON STANDARD SPIN 2 FIELD THEORY.

\vspace*{1cm}
A. L\'opez-Pinto{\footnote{alopezpinto@terra.es}}.

\end{center}}

It is usually accepted that General Relativity is the only consistent
theory which can be obtained starting from the linear Fiertz-Pauli lagrangian. It is the aim
of the present paper to study wether, under certain requirements, a different and 
consistent field theory can be found. These requirements will be the common ones
encountered in flat field theory: removal of the non physical degrees of freedom
and conservation of the energy and momentum currents determined from Noether's Theorem.
 It will be shown that imposing certain constraint (related to the elimination of the 
undesired components of the reducible representation) on the field manifold,
a consistent theory (at least to first order in nonlinearties) is achieved. The theory
obtained proceeding this way is characterized, to the lowest non linear order,  by certain parameter $\ep$. 
General Relativity's corresponding term is found to be the limit case of our non standard theory 
when $\ep \to 0$. So, $\ep$
measures at this level the size of the breaking of the  global symmetry appearing in 
General Relativity {\it{i.e.}} diff. invariance. It remains as open questions the matters
of the new theory's solvability to all orders and the appearence of it's quantized version.  

\ms

\section{\bf
 Introduction.}

\ms

\hs Since the times of Kraichnan [1], Gupta [2], Feynman [3] etc, the process for reobtaining
Einstein's General Relativity departig from an (inconsistent) spin 2 linear field theory
in the minkowskian space is widely known. We are aware that in the transit to a consistent 
theory many of the starting model's essential features are lost: the minkoskian space turns  
to be a pseudoriemannian
one, the interpretational meaning of coordinates changes, the global Lorentzian simmetry
gives way to a double fold symmetry (local Lorentz invarience plus global diffeomorphisms), and so on.
It is the purpose of the present paper to study the posibility of finding a nonlinear spin 2 consistent 
theory fulfilling the usual requirements encountered in flat non geometrical field theory, and to stablish which 
are the differences between our non standard construction and Einstein's geometrical one. We will not accept
any arbitrarity on the form of the expressions defining the theory and the determination of all the lagrangian terms
will be based on consistency or physical grounds.

\vs

   In section 2 a general review is made of the essential ingredients and requirements met in classical field theory.
Special emphasis will be done on gauge reduction, responsible of the non physical degrees of freedom
removal, and on the definition of the mathematical object which correctly describes the energy an momentum
currents ({\it{i.e.}} the energy momentum tensor, from here called e.m.t.). These two points will be shown to be crucial 
in our non standard construction. Section 3 is just the application of section 2 to a well known and
simple example: electrodynamics. Section 4 reviews the celebrated inconsistency of the linear spin 2 field theory. 
Mention is made of Deser's first order formalism outlook [4] and special prominence is given
to Feynman's approach [3] and to the consistency condition therein obtained. We also remind that it is commonly accepted
that General Relativity is the only proper theory fulfilling this condition. Section 5 deals with the topic
of trying to write a spin 2 field theory satisying section 2 demands. We show that there is certain
definition of the field manifold that solves the problems of  gauge reduction, Feynman's consistency
condition and is compatible (at least to first order in the nonlinearties) with the conservation criterion stated
in section 2.

\ms
\section{\bf
 Ingredients and requirements in a relativistic field theory.}

\ms

\hs We overview the essential ingredients and requirements encountered in classical field
theory. Points 2-i) to 2-iv) deal with the ingredients, while 2-v) and 2-vi) do with the 
requirements.\vs

 2-i)  Our departure arena is the minkowskian manifold $M_4$ and the corresponding coordinatizations
given by the assembly of inertial observers. As usual, those coordinatizations are related via
transformations of the Poincare group.
\vs

 2-ii)  We contemplate the case of $N$ particles, characterized through their $M_4$ trajectories 
$ {\{X_i\}}_{i=1,...,N}\;;\,{X_i}:\Re \rightarrow M_4$. We denote their components in a given
inertial system by $z^{\mu}_i\,(\lambda)$, where $\lambda$ is the line parameter, usually arc length. 
We also consider several quantities $(\psi_i ^1,...,\psi_i ^k)$ related to the i-th particle which 
are found to be outstanding for a given theory. For instance, when studying electrodynamics
we choose $\psi_i ^1
= m_i$ y $\psi_i ^2=q_i$, mass and charge of the i-th particle. Of course, the generalization
of these quantities to the case of finite distributions is immediate introducing the densities
$(\psi_i ^1 (p),...,\psi_i ^k (p))\;\; p\in M_4$.\vs

2-iii)  We also deal with a manifold $B$ of fields defined over $M_4$. The "points" of $B$
are the precise values of the fields: $ {\{B_j\}}_{j=1,...,Q}$, where $ {B_j}: M_4\rightarrow W_j$ and $W_j$ 
is some representation space. In the theories we shall consider, the fields $B_j$ transform
as Lorentz or Poincare representations. \vs

2-iv)   The system's dynamics is determined via the specification of the action functional $S$. 
It is generally understood that $S$ can be broken into three pieces, the particles', the field's 
and the interaction particles-field's term:

\ba S=S_{P} + S_{F} + S_{I}\nn \ea

As usual, the motion an field equations are obtained equating to zero the corresponding
"fixed" functional variations
\ba
\delta\vert_{fields fixed} \;\bigl( S\bigr)& \equiv\,\delta S_P +\delta\vert_{fields fixed} \;
\bigl(S_I\bigr)=\,0\nn \\
\delta\vert_{traj. fixed}\; \bigl(S\bigr) & \equiv\,\delta S_F +\delta\vert_{traj. fixed}\; 
\bigl(S_I \bigr)=\,0\,.\nn
\ea

2-v)    Both of the action's addends ($S_F$ and $S_I$) where the fields
$B_j$ appear are written as either line integrals or 4-integrals over $M_4$. So some of the field indices
are saturated with the ones belonging to the 4-vector $\dot z^\mu$ or with the derivative operator
$\p_\mu$. Therefore it is easly understood
that the natural representations fitting the action are tensorial ones. It is well known that 
these are neither Poincare nor Lorentz irreps. However
we usually want to deal with fundamental fields, which are labeled by definite Cassimir values (mass and
spin) and are found to be irreducible representations of the Poincare group. Therefore the mathematical
objects we handle when writting the action are carriers of superflous degrees of freedom. This drawback
is settled introducing some kind of "gauge reduction" {\it{i.e.}} certain internal transformation
 ${\{B_j\}}\rightarrow {\{B^{\prime}_j\}}$ leaving the action invariant:
 $S\,(B)\,=\,S(B^\prime)$. This invariance allows to remove the unwished degrees of freedom. As we shall 
later see, this is not the only acceptable way to proceed, and the gauge reduction condition can be performed
in different manners. Nevertheless we impose the following requirement on our theory. \vs

{\underline{Requirement 1}} - A classical field theory should be such that it enables
the removal of the undesired degrees of freedom. 

\vs

2-vi)   The underlying Poincare symmetry determines the existence of two essential objects related
to conserved quantities. Associated to the homogeneous part of the group we have
the angular momentum tensor, while the e.m.t. is the object which takes account of translational invariance.
We shall study how to define the e.m.t. and under what conditions it satisfies the conservation requirement.\vs

  For both free particles ($S=S_P$) and free fields ($S=S_F$) use can be done of Noether's Theorem
machinery in order to determine the corresponding conserved current. In first case we obtain
the particle's e.m.t.

\ba \tau_{ \mu\nu}=mc \,\delta^{(3)}\bigl({\vec r} -{\vec r}(x_0)\bigr)\, \dot z_\mu \dot z_\nu {{ds}\over{dx^0}}\ea

\nind where $\vec r (x_0)$ is the 3-trajectory of the mass $m$ particle and 
$\dot z_\mu$ it's 4-speed. 
Generalization to $N$ particles is achieved adding the label $i$ to $\{m,\vec r(x_0), z_\mu\};\quad 
i=1,...,N$; and to the discrete case introducing a matter distibution function. In the free particle 
case $\ddot z_\mu\,=\,0$ and the conservation condition is automatically
fulfilled

\ba  {\tau_{\mu\nu}}^{,\,\mu}=0\,.\nn \ea

\vs

  In the free field context we proceed similarly and, making use of Noether's Theorem, the following conserved
tensor is obtained  (see [5])

\ba T^{\mu\nu}=L^\mu_\alpha \,B^{\alpha ,\nu} -L_F\,\eta^{\mu\nu}-{f^{\mu\sigma\nu}}_{,\,\sigma}\,,
 \ea

  \nind where $B^\alpha $ are the fields, $L_F$ is the lagrangian field density and 

\ba L^\mu_\alpha \equiv {\partial L_F\over{\partial B^\alpha_{,\,\mu}}}\,\,, \nn\ea

\ba f^{\mu\sigma\rho}\equiv {1\over 2} \bigl(L^\mu_\alpha S^{\alpha\sigma\rho}_\beta+L^\sigma_
\alpha S^{\alpha\rho\mu}_\beta
-L^\rho_\alpha S^{\alpha\mu\sigma}_\beta\bigr)\,B^\be\,\,,\ea

\nind with $S^{\alpha\sigma\rho}_\beta$ the quantities related to the infinitesimal Poincare 
transformations of the fields via

\ba B^{\prime\alpha}(x^\prime)=\Biggl[\delta^\alpha_\beta+{1\over 2}\,S^{\alpha\mu\nu}_\beta 
\omega_{\mu\nu}\Biggr]\,B^\beta (x) \;\nn \ea

\nind and $\omega_{\mu\nu}$ the parameters of the infinitesimal homogeneous transformation.\vs

\vs
Makig use of the free field's equations is not hard to see that $T^{\mu\nu}$ identically satisfies

\ba { T^{\mu\nu}}_{,\,\mu}=0\,\nn\ea

  We agree on calling the object (2.2) the true energy momentum tensor of the fields. 
Due to historical reasons it is known in the literature as Belifante`s e.m.t. It is important
to draw attention to the fact that (2.2) is precisely the object obtained 
when aplying Noether's Theorem to the translational invariant $L_F$. It should 
also be emphasized that, despite what is quite often asserted, the last addend appearing in (2.2)
is not introduced {\it{ad hoc}} to symetrize the canonical e.m.t. It naturally
appears as a consequence of dealing with Noether's Theorem's machinery. As a fact 
${f^{\mu\sigma\nu}}_{,\,\sigma}$ takes into account the difference
$B^{\,\prime\be}(x^\prime)-B^\be(x)$ \vs {\it{i.e.}} the non scalar nature of the fields.

  We know that the quantities $T^{\mu\nu}$ are not observable. Nevertheless, the ones obtained through
the 3-integration 
 
\ba P^\nu =\int d\sigma_\mu T^{\mu\nu},\ea

\nind are by theirselves observable. Of course, $P^\nu$ is the energy momentum 4-vector of 
the free field.

\vs
    (2.4) suggests the following definition. We say that the expressions 
$H^{\mu\nu}$ and
$G^{\mu\nu}$, both written in terms of the field $B^\alpha$, belong to the same 1-orbit if

\ba \int d\sigma_\mu H^{\mu\nu}=\int d\sigma_\mu G^{\mu\nu}+A\;,\ea

\nind where $A$ is an arbitrary constant and it is understood that use has been made of the free
field's equations in the equality (2.5). Taking into account that the observable object is not 
$T^{\mu\nu}$ but $P^\nu$
we realize that, in the free field case,  $T^{\mu\nu}$ given by (2.2) is not the only acceptable object  
which describes energy momentum densities, but any other construction belonging to the same 1-orbit. 
For instance, the canonical e.m.t. $\buildrel \circ\over{{T}}^{\mu\nu}$, which is nothing but the part of 
$T^{\mu\nu}$ that does not deal with the non scalar nature of the fields {\it{i.e.}}

\ba{\buildrel \circ\over{{T}}}^{\mu\nu}=L^\mu_\alpha \,B^{\alpha ,\nu} -L_F\eta^{\mu\nu}\;,\nn\ea

 \nind belongs to the same 1-orbit as $T^{\mu\nu}$ and both are equally acceptable for describing the free field.
$T^{\mu\nu}$ is usually preferred to the canonical e.m.t. because the first one's construction
ensures  symmetry under the exchange  $\mu\leftrightarrow \nu$ and also because $T^{\mu\nu}$ is the object
which naturally appears imposing the condition ${\bar\delta}S=0$, where $\bar\delta$ stands for an infinitesimal 
Poincare transformation. It should be stressed that if no transit is made to the interaction case,
it is meaningless to distinguish between different elements in a 1-orbit, as far as (before
switching on the interaction!) they provide the same physical predictions. We will use the brackets
$\bigl\{\;\bigr\}_1$ to denote a given 1-orbit; for instance $\bigl\{T^{\mu\nu}\bigr\}_1$ is the 1-orbit
containing both $T^{\mu\nu}$ and $\buildrel \circ\over{{T}}^{\mu\nu}$.

\vs

   We now "switch on" the field-particle interaction {\it{i.e.}} we consider a theory described by the action
$S=S_P+S_f+S_I$. It is quite natural to expect from the new situation to satisfy energy-momentum conservation
through the equality 

  \ba \Bigl(\tau_{\mu\nu}+{\tilde T}_{\mu\nu}\Bigr)^{,\,\mu}=0\;, \ea

\nind where, in order to reobtain the free field predictions when the interaction is switched off,
${\tilde T}_{\mu\nu}\,\in\,\bigl\{T^{\mu\nu}\bigr\}_1$.

\vs

We will say that $H^{\mu\nu}$ and $G^{\mu\nu}$ belon to the same 2-orbit if (2.5) is fulfilled
in the new context, {\it{i.e.}} 
making use of the interaction field equations.\vs

Of course, and similarly to the free field case, if $\buildrel 1\over{{T}}_{\mu\nu}$
and $\buildrel 2\over{{T}}_{\mu\nu}$ are such that $\bigl\{\buildrel 1\over{{T}}_{\mu\nu}\bigr\}_2
=\bigl\{\buildrel 2\over{{T}}_{\mu\nu}\bigr\}_2$, both objects describe the same 4-momentum context
once the interaction is on.
The main difference lies in the fact that we now have a definite criterion to discriminate some
of the objects in a given 2-orbit, as far as the free field acceptable values are already known. So,we shall
request the object appearing in (2.6) to belong to the 1-orbit $\bigl\{T^{\mu\nu}\bigr\}_1$.\vs

   It is not hard to realize that $T^{\mu\nu}$ and $\buildrel \circ\over{{T}}^{\mu\nu}$
do not belong to the same 2-orbit (though they were in the same 1-orbit). As far as the relation
between the e.m.t. and the angular momentum tensor suggests symmetry of the first one, we will 
 {\it{a priori}} consider as an acceptable candidate to describe the e.m.t any object 
${\tilde T}_{\mu\nu}$ such that ${\tilde T}_{[\,\mu\nu\,]}\,=\,0$ and ${\tilde T}_{\mu\nu}\in
\bigl\{T_{\mu\nu}\bigr\}_1$.\vs

   Nevertheless the choice of the true e.m.t. $T^{\mu\nu}$ as the correct object ${\tilde T}^{\mu\nu}$
appearing in (2.6) lies not only on the satisfaction of the before mentioned conditions but also on
the verified fact that it is precisely this object the one which satisfies (2.6) in the well tested electromagnetic
case. So we request the theory to fulfill the following  \vs

    {\underline{Requirement 2}}. The action describing a realistic field theory should be such that

\ba \Bigl(\tau_{\mu\nu}+{ T}_{\mu\nu}\Bigr)^{,\,\mu}=0\;, \nn\ea

\nind where $\tau_{\mu\nu}$ and $T_{\mu\nu}$ are respectively given by (2.1) and (2.2).\vs

   Before endig this point we emphasize the following  \vs

a) Not all the actions satisfy Req.2. We will say that a theory is truly conservative
 if it fulfills Req.2. So Req.2 can be restated saying that our field theory sholud be
truly conservative.\vs

b) When dealing with a non truly conservative theory, the common procedure is to seek a symmettrical object 
$\Upsilon^{\mu\nu}(B_\alpha)$  which, via the field and motion equations satisfies 

\ba \bigl(\tau_{\mu\nu}+{\Upsilon}_{\mu\nu}\bigr)^{,\,\mu}=0\;.\nn\ea

\vs

     It should of course be checked that $\Upsilon^{\mu\nu}\,\in\,\bigl\{T^{\mu\nu}\bigr\}_1$.\vs

   Once reviewed the essential features of a flat field theory we aply them to a familiar case.

\ms

\section{\bf
 A simple example: spin 1 electrodynamics.}
\setcounter{equation}{0}

\ms 
\hs In order to illustrate last section we consider the well known case of electrodynamics. Regarding
points 2-i) to 2-iv), the elements involved are\vs

\nind -  the minkowskian manifold $M_4$;\vs

\nind - a particle of mass $m$, charge $e$ and 4-trajectory $z^\mu (s)$;\vs

\nind - the vectorial field $A^\mu$;\vs

\nind - the action functional

\ba S=S_P+S_I+S_F=\, -mc\int ds -{e\over c}\int A^\mu dz_\mu -{1\over 4}\int
F^{\mu\nu}
F_{\mu\nu}\,d^4x\;,\nn\\
\ea

\nind and the corresponding motion and field equations

\ba mc{\buildrel{..}\over z}^\mu = \,{e\over c}\,F^{\mu\nu}\,\dot z_\nu \ea  
 \ba{F^{\mu\nu}}_{,\mu}=  \,{1 \over c}\,j^\nu\;, \ea

\nind where $j^\nu$ is the 4-vector current $j^\nu\equiv \, e\,{{dz^\nu}\ /{d{x_0}}}\,\,
\delta^{(3)}(\vec r -\vec  r (x_0))$\vs

  Regarding 2-v), $A^\mu$ does not carry  an irrep of the transformation group but breaks into
direct sum of the $s=1$ irrep and the (undesired) $s=0$ one. According Req.1 we expect
the superflous degree of freedom to be somehow removed. Taking into account the functional dependence of
(3.1) it is not hard to see that equations (3.2) and (3.3) are invariant under the internal
transformation $A_\mu \to A_\mu +\,\phi_{,\mu} $, with $\phi$ an arbitrary scalar field with vanishing
divergence on the border of the integration domain of $S_I$. This invariance implies a degeneration
of the coupled system (3.2) + (3.3) responsible of the expected gauge reduction.\vs

  It should be remarked that passing from the free field case, characteized by the equations ${F^{\mu\nu}}_{\,,\mu}=0$ 
to the interaction context, where the corresponding equations are (3.2) + (3.3), the symmetry
underlying the gauge reduction does not break down. This is due to the fact that the
source of the field trivially fulfills the condition ${j^\mu}_{,\,\mu}=0$. Of course, this equality implies
no condition neither on the trajectory $z^\mu(s)$ nor on it's derivatives. This is one of the 
essential differences between the spin 1 and the spin 2 cases.\vs

   Regarding 2-vi) it is not hard to see that (3.1) describes a truly conservative theory
({\it{i.e.}} it satisfies Req.2). It can be checked that the corresponding true e.m.t. is

\ba T^{\mu\nu}={F^\mu}_\alpha\,F^{\alpha\nu}\,+{1\over 4}\eta^{\mu\nu}\,F^{\rho\epsilon}
\,F_{\rho\epsilon}\;,\nn\ea

\nind which is symmetrical and fulfills (2.5).\vs

  The relation connecting the true e.m.t. and the canonical one is 

\ba T^{\mu\nu}=\,\buildrel \circ\over{{T}}^{\mu\nu}+\,F^{\mu\sigma}\,{A^\nu}_{,\,\sigma}
\;.\nn \ea

 As expected $T^{\mu\nu}$ and $\buildrel \circ\over{{T}}^{\mu\nu}$
belong to the same 1-orbit but they are in different 2-orbits, as far as
$\int F^{\mu\sigma}\,{A^\nu}_{,\,\sigma}\,d\sigma_\mu$ vanishes when the fields $A_\mu$ 
follow the free field equations.\vs

  We finish quoting without demonstration the following statement: the form of the
 spin 1 linear field lagrangian appearing in (3.1) can be determined by either imposing Req.1 
or Req.2.

\ms

\section{\bf
 The spin 2 case. General Relativity.}
\setcounter{equation}{0}
\ms

\hs It is widely known that the path followed to obtain the commonly accepted
spin 2 field theory, {\it{i.e.}} Einstein's General Relativity (from here on G.R.), differs from the 
one sketched in section 2. The cornerstones for G.R. are covariance, the strong equivalence
principle and the requirement on the action of being second order in the fields' derivatives.
These conditions suffice to determine the Hilbert-Einstein's (from here on H-E) 
lagrangian and of course no reference
needs to be done to the points listed in section 2.\vs

  Anyway it is an old issue (see [1],[2],[3],[4]) to try to reobtain G.R. 
as the limit case of a flat field 
theory. We shall review at bird sight two of these constructions checking wether they fulfill the 
requirements stated in section 2.\vs

In this direction we start considering the symmetrical field $h_{\alpha\beta}$, which is the mathematical
carrier of the gravitational interaction (for a justification see [4] or [6]).
We call $H$ the manifold of the second order symmetrical tensors {\it{i.e.}}
$H\,=\,\bigl\{\,h_{\al\be}\;:\; h_{\,[\al\be ]}\,=\,0\,\bigr\}$.
Following section 2 we start with the minkowskian manifold $M_4$ and a particle of mass $m$ characterized
by its' 4-trajectory $z^\mu$. We also consider the action functional

\ba S_0=\, -mc\int ds -\lambda\int h^{\mu\nu}\tau_{\mu\nu}\,d^4x
+\int L_{F.P.}d^4x\;,\nn\ea

\nind where the appearence of the particle's e.m.t. $\tau_{\mu\nu}$ in the interaction term
is due to: i) the tensorial rank of the field, ii) the requirement of locality and
iii) the imposition that the highest degree of ${{d^n}\over{ds^n}}\,z^\mu\,(s)$ in the
equation of motion is given by $n=2$. $\lambda$ is the coupling constant. 
Concerning last term in $S_O$, $L_{F.P.}$ is the Fiertz-Pauli
lagrangian. It can be obtained imposing Req.1 to the free field case,
taking into account that $h_{\alpha\beta}$ carries undesired degrees of freedom different 
from the spin 2 ones.
They can be removed using the condition 
$\Bigl({{\delta L_{F.P.}}\over{\delta h_{\alpha\beta}}}\Bigr)_{,\alpha}\equiv 0$.
This reduces in  4 the number of linearly independent free field equations and implies the 
invariance of these equations under the internal transformation $h_{\alpha\beta}
\to h_{\alpha\beta}+\chi _{\{\alpha,\beta\}}$, with 
$\chi_\alpha$ an arbitrary vectorial field. Imposing the before mentioned 
condition it is not hard to see that

\ba L_{F.P.}\equiv {1\over 2}\,h_{\epsilon\gamma ,\alpha}h^{\epsilon\gamma ,\alpha}
-{h_{\epsilon\gamma}}^{,\epsilon} {h^{\gamma\beta}}_{,\beta}
+{h_{\epsilon}}^{\epsilon ,\beta} {h_{\alpha\beta}}^{,\alpha}
-{1 \over 2}\,{h_{\epsilon}}^{\epsilon ,\alpha} {h^{\gamma}}_{\gamma,\alpha}\;.\nn
\ea

\vs

   Problems come when the interaction is switched on, as the new field equations

\ba -\lambda\tau_{\mu\nu}+{{\delta L_{F.P.}}\over{\delta h^{\mu\nu}}}=0\nn\ea

\nind leads to the pointless condition 

\ba\ {\tau^{\mu\nu}}_{,\,\mu}=0\,;\ea

\nind senseless as far as $\ddot z ^\nu\,\not=
\,0$. We try to recuperate coherence rewriting the action as

\ba S=\, -mc\int ds -\lambda\int h^{\mu\nu}\tau_{\mu\nu}\,d^4x
+\int \bigl( L_{F.P.}+\Delta L\bigr)d^4x\; \ea

\nind whre $\Delta L$ are new terms appearing in the field lagrangian; terms of order
higher than 2 in $h_{\alpha\beta}$ and so responsible of the nonlinearty of the theory. 
Roughly speaking, the object of the before mentioned references ([1],[2],[3],[4])
is to determine, once given
suitable consistency conditions, how do $\Delta L$ looks like. Of course the expected result, 
{\it{i.e.}} G.R. is always obtained. The difference between the different approaches 
found in the literature lay on the particular
procedure followed and (as denounced by Weinberg in [6]) in the amount of arbitrarity in the election 
of the corresponding "specific lagrangians".
\vs

 We will start reviewing [4]. It was published quite later than the rest of
the before cited papers, but it has the advantage of being particularly direct. Deser
considered a first order lagrangian $L_D \,\bigl(\varphi^{\mu\nu},
\Gamma^\alpha_{\beta\epsilon}\bigr)$, where $\varphi^{\mu\nu}$ is a field
related to the deviation from flatness of the covariant metric density
and $\Gamma^\alpha_{\beta\epsilon}$ is an (independent of $\varphi^{\mu\nu}$) simmetric
connection. $L_D$ is such that the "$h$" and "$\Gamma$" field equations make it
 equivalent (modulo a field redefinition) to the second order lagrangian $L_{F.P.}$.\vs

  We already know that the inconsistency (4.1) is mended
adding  source terms $^{(j)}\theta_{\mu\nu}$, which are
increasing order energy momentum contributions due to the self interaction of the field, to the field
equations. This is an alternative point of view to the $\Delta L$ addition in (4.2). Following 
this line Deser added a source term to the field equations which was somehow related to the e.m.t. of the original
lagrangian. The essential point in the construction is that the action giving way (via variational derivation) to the
modified equations can be rewritten as the {\it{a la}} 
Palatini version of H-E action. So, proceenimg this way we have succeed the program 
of passing from the linear form to the
non linear equations which take into account all the source terms needed to fix inconsistency; 
and these source precisely reproduce the energy-momentum contribution of the complete
field lagrangian. Furtermore the
final result is equivalent to G.R.
No iteration is needed as far as as in first order formalism G.R. is third order in the
chosen variables. 
\vs

 Nevertheless we find several uneasy features in Deser's construction. First of all
the energy momentum conservation condition used as consistency condition seems to fade away when translated 
to the natural second 
order formalism: according [4] obtention via the action's functional derivative 
of an expression equivalent to Belifante's e.m.t. is the cornerstone of the construction. 
Direct calculation shows that
this is not the case in second order formalism, as far as the condition 

\bq T_{\alpha\beta}={{\delta(L_{F.P.}+\Delta L)}\over{\delta h^{\alpha\beta}}}\nn\eq
  
\noindent implies that the second order version of $T^{\mu\nu}$ {\it{i.e.}}
$T^{(2)}_{\mu\nu}$ can be written as ${{\delta L_{(3)}}\over{\delta h^{\mu\nu}}}$ 
(see (4.4) for the meaning of $L_{(3)}$).
It is not hard to check that it is not possible to adjust the coeficients of $L_{(3)}$ in order to satisfy (4.3).
\vs

  Furthermore Deser's construction 
is not exclusive in the sense that it does not rule out the possibility of (using second order
formalism) finding a non linear theory with linear part given by the Fiertz-Pauli piece but, when going 
to higher orders, different from G.R.
 Another uneasy feature of [4] is that it does not provide a 
definite criterion (like the on given in eq. (6.2.3) of [3]) to "discriminate" wether a tempative theory is acceptable
or not. Besides, 
it is quite hard to determine if the procedure described in [4] 
satisfies the program sketched in section 2. As a fact, the use of first order variables darkens 
the study of Req.1 and (as already mentioned) Req.2. \vs

The approach followed by Feynman [3] is quite tiresome, for use is made of second order formalism which 
is the natural language of the problem, but requires to take into account the contributions of
an infinite series. Lengthy calculations become more and more horrible when going
to higher orders. The nice side of the approach is that it provides a necessary
condition on $\Delta L$ based on the simultaneous fulfillment of the field and motion equations.

Starting from the general action (4.2), we break $\Delta L$ into the sum 

\bq \Delta L=\sum_{j=3}^\infty L_{(j)}.\nn\eq

\vs 

In order to simplify the notation we schematicaly write 

\bq L_{(j)}=\sum_{n=1}^q a^j_n\,\Bigl( h^{j-2}\p h\p h\Bigr)_n\;,\eq

\nind where $\Bigl\{\Bigl( h^{j-2}\p h\p h\Bigr)_n\Bigr\}_{n=1}^q$ stands for the set of $q$ linearly
independent (modulo an additive 4-divergence) terms of order $(j-2)$ in the non derivative $h^{\prime s}$
and order 2 in the derivative ones. $\bigr\{a^j_n\bigl\}$ are certain constants. To clarify the notation 
we consider the  Fiertz-Pauli lagrangian ({\it{i.e.}} the $L_{(2)}$ case) and define

\ba a^2_1\equiv{1\over 2}\quad , \quad\Bigl( \p h\p h\Bigr)_1\equiv 
h_{\epsilon\gamma ,\alpha}h^{\epsilon\gamma ,\alpha}\nn &\\ 
a^2_2\equiv-1\quad , \quad \Bigl( \p h\p h\Bigr)_2\equiv
{h_{\epsilon\gamma}}^{,\epsilon} {h^{\gamma\beta}}_{,\beta}\nn & \\ 
a^2_3\equiv1\quad , \quad \Bigl( \p h\p h\Bigr)_3\equiv
{h_{\epsilon}}^{\epsilon ,\beta} {h_{\alpha\beta}}^{,\alpha}\nn &\\ 
a^2_4\equiv-{1\over 2}\quad , \quad \Bigl( \p h\p h\Bigr)_4\equiv
{h_{\epsilon}}^{\epsilon ,\alpha} {h^{\gamma}}_{\gamma,\alpha}\nn
 & \,.
\ea
\vs

Returning to (4.2) it is straightforward to realize that corresponding motion and field equations are

\ba&\bigl(\eta^{\mu\nu}+2\lambda \,h^{\mu\nu}\bigr)\,\ddot z_\mu
+2\lambda {h^{\mu\nu}}_{,\rho}\,\dot z^\rho\dot z_\mu 
-\lambda h^{\mu\tau,\nu}\,\dot z_\mu\dot z\tau=0\ea
 
\ba -\lambda \tau_{\mu\nu}+ {{\delta L_{F}}\over{\delta h^{\mu\nu}}}=0\;,\ea

\nind with $L_F\equiv L_{F.P.}+\Delta L$. It is not hard to check that (4.6) leads to

\bq g_{\mu\lambda}\,{\tau^{\rho\mu}}_{,\rho}=-[\mu\rho ,\lambda ]\,\,\tau^{\mu\rho}\;,\eq

\nind where used has been made of the following notation

\ba &g_{\mu\nu}\equiv \eta_{\mu\nu}+2\lambda \,h_{\mu\nu}\,,\\
\,&\nn\\ 
&[\mu\rho ,\nu ]\equiv \lambda\,\,\bigl[h_{\mu\nu,\rho}
+h_{\rho\nu ,\mu}-h_{\mu\rho ,\nu}\bigr]\;.\nn\ea

 Using (4.8) and the field equation (4.7) we infer the following consistency condition:

\vs
{\hspace*{4cm}}\ul{{\hspace*{7cm}}}
\vs

 {\bf{\underline{ Condition A.}}}   $L_F$ 
appearing in the action (4.2) of consistent spin 2 field theory necesarily fulfills the condition 

\ba g_{\mu\lambda}\,\Biggl({{\delta L_F}\over{\delta h_{\rho\mu}}}\Biggr)_{,\rho}
=-\bigl[\mu\rho ,\lambda \bigr]\,\, {{\delta L_F}\over{\delta h_{\rho\mu}}}\;\quad;\qquad 
 h_{\mu\rho}&\in H \;.  \ea
\nind where $L_F$ is the complete field lagrangian
\vs

{\hspace*{4cm}}\ul{{\hspace*{7cm}}}
\vs

    From last expression it is possible to recusively infer
the form of $L_{(j)}$ using $L_{(k)}\;,\,k < j$. Starting from $L_{F.P.}$ we use the second
order version of (4.10) to determine $L_{(3)}$ through

\ba \eta_{\mu\lambda}\,\Biggl({{\delta L_{3}}\over{\delta h_{\rho\mu}}}\Biggr)_{,\rho}
=-\bigl[\mu\rho ,\lambda \bigr]\, {{\delta L_{F.P}}\over{\delta h_{\rho\mu}}}.
 \ea

\vs

   Once obtained $L_{(3)}$ we can use the  third order version of (4.10) to determine 
$L_{(4)}$, and so on.\vs

  Proceeding this way we obtain (see Appendix A)

\ba L_{(3)}=\lambda\;\Biggl(-2\,[1]-2\,[2]+2\,[4]+4\,[5]-[6]+{1\over 2}\,[7]-\nn\\-[8]-[9]+2\,[10]-3\,[11]
+[12]+[13]-{1\over 2}\,[14]\;\Biggr)\;,\ea

\nind where the value of $[j]\;;\;j=1,...,14$ is given in Appendix A.\vs

  We can make use of (4.9); identify $g_{\mu\nu}$ with a metric tensor and write H-E
lagrangian as a function of $h_{\mu\nu}$. Working this way we check that, not only
the second order term obtained reproduces $-2\,\lambda^2$ times $L_{F.P}$, but the third 
order one is precisely $-2\,\lambda^2\,L_{(3)}$.
Instead of going to higher and higher orders in (4.10),  Feynman observed that condition A 
can be rewritten as an invariance condition of $L_F$ under infinitesimal transformations
of $h_{\alpha\beta}$. This is precisely the invariance underlying H-E lagrangian and 
so G.R. is a solution of (4.10) to all orders. So, imposing Condition A to a standard
field theory the machinery of
pseudoriemannian manifolds and differential geometry is recovered. \vs

Several remarks should be done:\vs

  a) Not only G.R. fulfills Condition A but, as far as we know, it is 
the only thory of the form (4.5) + (4.10) which follows it .\vs

  b) In the "intergated to all orders" theory, {\it{i.e.}} G.R., the gauge reduction given by 
invariance under $h_{\alpha\beta}\to
h_{\alpha\beta}+\chi_{\{\al,\beta\}}$, with finite $\chi_\beta$, is no longer valid.
Instead, an infinitesimal version underlies the new context. It is well known
that these transformations reproduce infinitesimal general diffeomorphisms. 
Initially Einstein elucidated this symmetry as the one due to the change 
in the observer's state of motion. As it will be demonstrated in [7],
this interpretation is not attainable. Anyway, the passing from the finite to the 
infinitesimal symmetry is closely related to the passage from flat to curve scenario. \vs

c) The before described reobtention of G.R.has been achieved without necessity
of invoking any restriction on the form of the field e.m.t. Only {\it{a posteriori}} and once
determined the aspect of $\Delta L$, we tackle the problem of studying the corresponding
e.m.t. Anyway it is clear that the true e.m.t. is not an accetable object to describe the
energy-momentum density, as far as it would imply 
\ba {T^{\mu\nu}}_{,\mu}+{1\over \lambda }\,
\Biggl({{\delta L_F}\over{\delta h_{\mu\nu}}}\Biggr)_{,\mu}=0\;,\nn\ea

\nind which is a condition over $\Delta L$ that G.R. does not satisfy. Actually
it is an unattainable condition over {\it{any}} $\Delta L$, (see Appendix B). 
Looking for coherence with section 2 we might seek a symmetrical object $\Theta^{\mu\nu}$,
which belongs to  $\bigl\{T^{\mu\nu}\bigr\}_1$ and 
fulfills  

\ba {\Theta^{\mu\nu}}_{,\mu}+{1\over \lambda }\,
\Biggl({{\delta L_F}\over{\delta h_{\mu\nu}}}\Biggr)_{,\mu}=0\,.\,\nn\ea

 Nevertheless this is not the common procedure. It is well known that  G.R. bears a 
new frame where there is no room for section 2
requirements. As a fact, in the geometrical scenario use is costumarily done of
 Weyl's e.m.t.:

  \ba T_W^{\mu\nu}=\,\gamma\,\Biggl({{\delta L_G}\over{\delta g_{\mu\nu}}}\Biggr),\nn\ea

\nind where $L_G$ is the geometrical field lagrangian (usually H-E) 
and $\gamma$ is afactor dependent of the lagrangian normalization
and chosen in such a way that $T_W^{\mu\nu}$ identically fulfills (2.5). 
Anyway, the subject of the e.m.t. in G.R.
is complicated, as far as there are several candidates deppendig, for instance, if
a field theoretical approach to the problem or a geometric 
one is used. There is a wide literature
on the subject (see [8]).
\ms

\section{\bf
 Non standard nonlinear spin 2 theory.}
\setcounter{equation}{0}
\ms

  \hs With all we have seen  and taking into account Condition A, it might seem nonsensical to seek a 
spin 2 theory based on different grounds, as far as (4.10) is a necessary condition on any consistent theory. 
Anyway we will temporary ignore it and deal with the question of determining
$\Delta L$ appearing in (4.2) following the path sketched in section 2.\vs

  We already mentioned that both Req.1 or Req.2 suffice to find the form
of the spin 1 field lagrangian. We know study wether something similar happens in a non linear spin 2
context. Unfortunately the gauge reduction condition Req.1 has litttle to say about the non linear action.
It was found to be enough to obtain the free lagrangian  $L_{F.P.}$, but apparently it  
has no influence on $\Delta L$. If we try to build a parallel condition to the one used to determine
$L_{F.P.}$ we observe that $\Biggl({{\delta \Delta L}\over{\delta h_{\alpha\beta}}}\Biggr)_{,\,\alpha}= 0$ is unattainable.
 Besides the before mentioned condition would not fix the
inconsistency (4.1). Furthermore the new nonlinear action does not
seem to be invariant under the old finite transformation $h_{\alpha\beta}\to
h_{\alpha\beta}+\chi_{\{\al,\beta\}}$. So, in the search of a non linear theory, it is by no means
evident to find a clue from gauge reduction.
\vs

  We next consider the consequences of imposing Req.2. Together with the field equation (4.7), Req.2 suggests
the identification 

\ba  T^{\mu\nu} \buildrel{?}\over{=} \,{{-1}\over \lambda}\,\Biggl({{\delta L_{F.P.}}\over {\delta h_{\mu\nu}}}\,+\, 
{{\delta \Delta L}\over {\delta h_{\mu\nu}}}\Biggr)\; \ea

  \nind where the symbol $\buildrel{?}\over{=}$ means that in $L.H.S.\,\buildrel{?}\over{=}
\,R.H.S.$, it is necessary to check the existence of certain value of the coefficients
$\bigl\{a^j_n\bigr\}$ such that the equaliyt $L.H.S.\,=\,R.H.S.$ holds. \vs

We rewrite (5.1) to second order as 

\ba T^{(2)}_{\mu\nu} \buildrel{?}\over{=} \,{{-1}\over \lambda}\,\Biggl({{\delta 
L_{3}}\over 
{\delta h^{\mu\nu}}}\Biggr)\;,\ea

\nind where $T^{(2)}_{\mu\nu}$ is built using the the construction described in 2-vi), 
introducing the value of $L_{F.P.}$.
It is easy to check that the coefficients in $L_3$ cannot be adjusted in such a way that (5.2) holds.
\vs

  Condition (5.1) is too restrictive. In order to ensure the true conservativity of the theory
it is enough to demand the less stringent condition 

\ba  \Bigl(T^{\mu\nu}\Bigr)_{,\,\mu} \buildrel{?}\over{=} \,{{-1}\over \lambda}\,\Biggl(
{{\delta \Delta L}\over {\delta h_{\mu\nu}}}\Biggr)_{,\,\mu}\;,\ea

\nind  which can be written to lowest order as 

\ba -\lambda\,\Bigl(T^{(2)}_{\mu\nu}\Bigr)^{,\,\mu} \buildrel{?}\over{=} \,
\Biggl({{\delta L_{3}}\over 
{\delta h^{\mu\nu}}}
\Biggr)^{,\,\mu}\;.\ea

\vs

  (4.11) and (5.4) have similar structures. Both reproduce a system of linear equations
with many more equations than unknowns. The main difference lies in the fact that (4.11) is
solvable while (5.4) is not (see Appendix B). So we arrive to the conclusion that the structure
(5.3) is incompatible.  \vs

   Nevertheless it is possible to find a consistent way to handle the problem. In this direction
we consider the gauge reduction requirement. Instead of starting with an action invariant
under certain 4-parameter internal transformation we shall assume
that the field manifold $H$ is somehow reduced to a smaller manifold $H^*$ which is nothing but the 
restriction of $H$ to the (non algebraic) condition $\varphi_\nu(h_{\alpha\beta})=0$; {\it{i.e.}}

\ba H^*=\bigl\{h_{\alpha\beta}\in H: \varphi_\nu(h_{\alpha\beta})=0\quad;\;\nu =0,1,2,3\,
\Bigr\}\;.\nn\ea

\vs

  This is an alternative and admissible way to impose the removal of the undesired degrees
of freedom. We shall implement this condition imposing the requirement that the fields
appearing in the dynamical equations are restricted to belong to $H^*${\footnote {Actually
the condition  $\varphi_\nu(h_{\alpha\beta})=0$ on $H$ should be imposed prior to the  determination
of the dynamical equations given by the functional variation $\de$. Nevertheless the suggested
procedure demonstrates to be "good enough" when the reduction of the manifold is too complicated. 
To simplify we consider an easy example. 
Imagine we start from a 
kynematical manifold $\bigl[\omega_1,\omega_2\bigr]$, where $\omega_1$ and
$\omega_2$ are scalar fields,  which is reduced, via the condition 
$\varphi\,(\omega_1,\omega_2)=0$. We shall assume that we know how to invert the before mentioned condition and
write $\om_2=\om_2(\om_1)$.
So we introduce the  reduced manifold $\bigl[\omega_1\bigr]=\bigl[\omega_1,\omega_2(\om_1)\bigr]$. 
We define the new action
using the old one as $S\,\bigl(\omega_1\bigr) \equiv S\,\bigl(\omega_1,\omega_2(\om_1)\bigr)$. Our {\it{real}}
dynamical problem is to determine the solutions of $\;{{\de\,S\,\bigl(\omega_1\bigr)}\over{\de\,
\om_1}}=0$. If
we have problems to deal with this equation, {\it{i.e.}} if it is not trivial
to invert $\varphi\,(\omega_1,\omega_2)=0$ we observe that

\ba \;{{\de\,S\,\bigl(\omega_1\bigr)}\over{\de\,\om_1}}=
\,{{\de\,S\,\bigl(\omega_1,\om_2\bigr)}\over{\de\,\om_1}}\Big\vert_{\om_2=\om_2(\om_1)}\,+\;
\,{{\de\,S\,\bigl(\omega_1,\om_2\bigr)}\over{\de\,\om_2}}\Big\vert_{\om_2=\om_2(\om_1)}\;
{{\partial\,\om_2}\over{\partial\,\om_1}}\;,\nn\ea
\nind so, the simultaneous imposition of 
$\,{{\de\,S\,\bigl(\omega_1,\om_2\bigr)}\over{\de\,\om_1}}\Big\vert_{\om_2=\om_2(\om_1)}\,=0$ and
$\,{{\de\,S\,\bigl(\omega_1,\om_2\bigr)}\over{\de\,\om_2}}\Big\vert_{\om_2=\om_2(\om_1)}\;=0$ is a sufficient 
(though not necessary!) condition for a solution of 
$\;{{\de\,S\,\bigl(\omega_1\bigr)}\over{\de\,\om_1}}=0$. }}\vs

  In this direction we suppose that $\varphi_\nu(h_{\alpha\beta})$ has the following form

\ba \varphi_\nu(h_{\alpha\beta})\equiv
\Bigl( \eta_{\mu\nu}+2\lambda h_{\mu\nu}\Bigr)
\,\Biggl({{\delta L_F}\over{\delta h^{\rho\mu}}}\Biggr)_{,\rho}
+\bigl[\mu\rho ,\nu \bigr]\, {{\delta L_F}\over{\delta h^{\rho\mu}}}\;,\ea

 \nind and hence $H^*$ is the manifold of the symmetrical tensors $h_{\alpha\beta}$ fulfilling
$\varphi_\nu(h_{\alpha\beta})=0$, where $\varphi_\nu(h_{\alpha\beta})$ is given by (5.5).\vs

  It is important to emphasize the differences between (4.10) and our condition $\varphi_\nu
(h_{\alpha\beta})=0$.  (4.10) was used to determine the value of the coeficients
$\{a^j_n\}$  satisfying this requirement. In this way we arrived to the value of the parameters
describing Hilbert-Einstein's lagrangian. This procedure implies no previous condition
over the fields $h_{\alpha\beta}$. \vs

     Nevertheles the condition which defines $H^*$ {\it{i.e.}} $\varphi_\nu(h_{\alpha\beta})=0$
is not a condition over the lagrangian coeficients but over the own kinematically acceptable
fields $h_{\alpha\beta}$. It is important to realize that the coefficients which appear
in the H-E lagrangian are not accepatable for our gauge reduction condition, as far 
as they identically cancel (5.5) and so imply no condition over the fields $h_{\alpha\beta}$.\vs

   This approach simultaneously gives solution to the superflous degrees of freedom
removal and to the consistency problem before stated. Nevertheless if we want to complete the
program outlined in Section 2 we still have to deal with Req.2. In this direction we impose
the

\vs
{\hspace*{4cm}}\ul{{\hspace*{7cm}}}
\vs

{\bf{\ul{Condition B.}}} A theory consistent with the requirements developed in section 2
describing a spin 2 field is determined by the condition

\ba -\lambda\, \Bigl(T^{\mu\nu}\Bigr)_{,\,\mu} = \,\Biggl(
{{\delta \Delta L}\over {\delta h_{\mu\nu}}}\Biggr)_{,\,\mu}\qquad;\qquad 
h_{\alpha\beta}\in H^*\;,\ea

\nind where $T^{\mu\nu}$ depends on $\Delta L$ through (2.2) and $H^*$ is defined via (5.5).

\vs
{\hspace*{4cm}}\ul{{\hspace*{7cm}}}
\vs

   We know study the consistency of Condition B. In this direction we consider the
 lowest order version of (5.6)

\ba  -\lambda\,\Bigl(T^{(2)}_{\mu\nu}\Bigr)^{,\,\mu} = \,\Biggl(
{{\delta  L^{\scriptscriptstyle{N}}_{(3)}}\over {\delta h^{\mu\nu}}}\Biggr)^{,\,\mu}\qquad;\qquad 
h_{\alpha\beta}\in H^*\;,\ea

\nind where we have introduced the label $N$ (of nonstandard) to distinguish the lagrangian
fulfilling (4.10) from the one given by (5.6).
\vs

We have already seen that (5.4) is incompatible. Nevertheless, taking into account
the condition $h_{\alpha\beta}\in H^*$, one equation can be eliminated from the total system.
In other words, a new
system of equations is obtained writing one of the vectorial terms 
 appearing in $\varphi_\nu \,\bigl(h_{\al\be}\bigr)=0$
as afunction of the rest and substititing it next in (5.4). Surprisingly, the system obtained
proceeding this way is 
compatible, (see Appendix B). As a fact there is a degeneracy of  order 1 in it`s solution. It is not
hard to check that our third order solution can be parametrized as

\ba L^{\scriptscriptstyle{N}}_{(3)}=\lambda\;\Biggr((-2+\epsilon)\,[1]
+(-2+\epsilon)\,[2]+
(2-\epsilon)
\,[4]+4\,[5] 
+\Biggl(-1-{1\over 2}\ep\Biggr)[6]+\nn\\
+\Biggl({1\over 2}+{1\over 4}\epsilon\Biggr)\,[7]
+\Biggl(-1-{1\over 2}\epsilon\Biggr)[8]
+{\Biggl(-1+{1\over 2}\epsilon\Biggr)}[9]
+{(2-\epsilon)}\,[10]+\nn\\
+\Biggl(-3-{1\over 2}\epsilon\Biggr)\,[11]
+\Biggl(1+{1\over 2}\epsilon\Biggr)[12]
+\Biggl(1+{1\over 2}\epsilon\Biggr)[13]
+\Biggl(-{1\over 2}-{1 \over4}\epsilon\Biggr)\,[14]\;\Biggl)\;,\nn\\
\qquad\qquad\ea

\nind where $\ep$ is a real number.

\vs  The resolubility of the new system of equations means that Condition B can be 
implemented at least to first order in the nonlinear contribution ${ \Delta L}$. It
opens the horizon for the construction of a spin 2 theory written following the path
indicated in section 2.

 \vs

We now remark several important points.\vs

   a) (5.8) tends to the G.R. term (4.12) when $\epsilon\to 0$. It should be 
emphasized that G.R. can by no means be a particular solution of (5.6), as far as in Einstein's
theory  $\varphi_\nu(h_{\alpha\beta})$ identically vanishes and there is no reduction from
$H$ to $H^*$. So, solution (5.8) is valid 
$\forall\epsilon\in \bigl\{(-\infty,0)\,\bigcup\,(0,\infty)\bigr\}$\vs

b) A look at (5.8) shows that, to the given order, $\ep$ breaks G.R.'s global symmetry {\it{i.e.}} diff.
invariance. The absolute value of $\ep$ measures the size of this breaking.\vs

c) The fact that G.R. gives the correct predictions for the classical tests suggests that
 the absolute value of $\epsilon$ is small.\vs

d) If we write (5.6) to fourth order and find the system to be solvable (which for the time
being is only a hypothesis) we expect to find something like 

\ba L^{\scriptscriptstyle{N}}_{(4)}\,=\,
L^{\scriptscriptstyle{N}}_{(4)}\Biggl(\Bigl\{a_n ^4\Bigr\}^q _{n=1}\,,\;\ep\,,\;
\Bigl\{\ep^\prime _p\Bigr\}_{p=1}^r\Biggr)\nn\ea

where $a^4_n\in\Re$ are $q$ fixed constants, $\ep$ is the indeterminacy appearing
in $L^{\scriptscriptstyle{N}}_{(3)}$ and $\ep^\prime_p$ are $r$ free parameters responsible of the r-indeterminacy
present in the fourth order version of (5.6). Of course $0\leq r \leq q$.\vs

e) According to the footnote in page 15, the dynamical problem can be treated equating to zero
the functional derivative of the lagrangian fulfilling Condition B with respect to an arbitrary
$h_{\al\be}$ and imposing as a simultaneous equation $\varphi_\nu(h_{\alpha\beta})=0$.\vs 

f) It remains unknown wether the infinite number of systems involved in (5.6) are 
solvable. Going to higher and higher order does not ensure the integrability
to all orders. It would be necessary to find a qualitative argumento to determine it.\vs

g) At this point the sensible reader might be wondering what is the necessity for
"it all". We have already got G.R., a familiar theory which fulfills condition A to all orders 
and provides correct prediction for the classical tests. Then, why complicate life with
a tentative theory of dubious behavior when going to higher orders?. We find several good
reasons. First of all the new theory might introduce new features for quantization.
 Furthermore it shall be demonstarted in [7]  that intrinsecal diff. invariance entails unavoidable
complications. A general solution  fulfilling Condition B is not diff. invariant, but for small 
values of $\ep$, gives good predictions. So it keeps away from the undesired 
difficulties and maintains agreement with experiments. Besides, if condition B is integrable, we
have at our disposal a consistent spin 2 theory following the usual field theory requirements;
characterized by a well defined value of the e.m.t. (actually by the true e.m.t. introduced in
section 2).

\ms
\section{\bf Conclusions.}
\ms

\hs We have seen that, starting from the second order linear Fiertz-Pauli lagrangian, there seems
to be two roads which, going to a non linear field self interaction, avoid inconsistency. The first one 
is the well known Condition A, giving way to Einstein's G.R. The second is condition B which we have
proved to be integrable to third order in the field lagrangian. The corresponding term depends on 
certain free parameter which shows the deviation from the diff. invariance solution. It remains as an open question
wether a theory fulfilling condition B is feasible to all orders or not, and which are the quantum 
implications of the (hypothetical) complete field lagrangian.

\ms

\section{\bf Appendices.}
\setcounter{section}{1}
\setcounter{equation}{0}
\renewcommand{\thesection}{\Alph{section}}

\ms

{\large {\bf{A.}} {\ul{Some calculations concerning Condition A.}}}

\ms
Following (4.4) we write 

\ba L_{(3)}=\sum_{n=1}^q a^3_n\,\Bigl( h\,\p h\p h\Bigr)_n\;.\nn\ea

\vs

 It is easy to check that in this case $n=14$, {\it{i.e.}} we can choose $14$ 
$\Bigl( h\,\p h\p h\Bigr)$ linearly independent terms (modulo an additive 4-divergence) 
in such a away that the rest of the terms are written as linear combination
(modulo an additive 4-divergence) of the former $14$. In order to simplfy
the notation we write

\ba x_n \equiv\,a^3_n\;\qquad [n] \equiv \,\Bigl( h\,\p h\p h\Bigr)_n\;\qquad n=1,...,14\;,\nn\ea

\nind {\it{i.e.}}

\ba L_{(3)}=\,\sum_{n=1}^{14} x_n\,[n]\;.\nn\ea

\vs

We can choose\footnote {There are some algebraic errors in ref. [3] probably due to a misstranscription
of Feyman's lectures (though the conclussions are absolutely correct!). First of all the minimum number 
of elements of the form $h\p h\p h$ needed
to write $L_{(3)}$ is not 18 but 14. There is also an error in eq.(6.1.13). In order to check it, 
suffices to realize that (6.1.13) does not reproduce the coresponding term in the H-E lagrangian.}

\ba & [1]\,  \equiv\,h_{\beta\delta}\,h^{\alpha\beta,\gamma}\,{{h_\alpha}^\delta}_{,\gamma}\;;\;
[2]\,\equiv\,b^\beta\,h_{\alpha\gamma ,\beta}\,h^{\alpha\gamma}\;;\;
[3]\,\equiv\,b^\alpha\,b^\beta\,h_{\alpha\beta}\;;\;
 [4]\,\equiv\,h_{\alpha\beta ,\gamma}\,h^{\gamma\epsilon ,\beta}\,{h_\epsilon}^\alpha \;;\nn\\
 & [5]\,  \equiv\,h_{\alpha\gamma ,\beta}\,h^{\epsilon\alpha ,\gamma}\,{h^\beta}_\epsilon \;;\;
[6]\,\equiv\,h_{\alpha\gamma,\beta}\,h^{\alpha\gamma ,\epsilon}\,{h^\beta}_\epsilon \;;\;
[7]\,\equiv\,h\,h_{\epsilon\beta,\alpha}\,h^{\epsilon\beta ,\alpha} \;;\;
[8]\,\equiv\,h\,b_\epsilon\,b^\epsilon \;;\nn \\
 &  [9]\,\equiv\,c^\alpha\,h_{\alpha\epsilon,\beta}\,h^{\epsilon\beta}\;;\;
[10]\,\equiv\,c_\alpha\,h^{\beta\epsilon,\alpha}\,h_{\epsilon\beta}\;;\;
[11]\,\equiv\,h_{\epsilon\alpha}\,c^\alpha\,b^\epsilon \;;\;
[12]\,\equiv\,h\,c_\alpha\,b^\alpha \;;\nn\\
& [13]\,\equiv\,h_{\alpha\beta}\,c^\alpha\,c^\beta \;;\;
[14]\,\equiv\,h\,c_\alpha\,c^\alpha\;, \qquad\qquad\nn
\ea

\nind where

\ba b_\alpha\,\equiv\,{h^\epsilon}_{\alpha ,\epsilon}\;;\;
c_\alpha\,\equiv\,h_{,\alpha} \;;\; h\,\equiv\,{h_\epsilon}^\epsilon \nn\;.\nn\ea

\vs

  Using the notation

\ba\ \, & \,(1)_\be \,\equiv\,h_{\ep\be ,\al}\,\B h^{\al\ep}\;;\;
(2)_\be\,\equiv\,h^{\ep\al}\,\B h_{\be\ep ,\al}\;;\;
(3)_\be\,\equiv\,h^{\al\ep}\B h_{\al\ep ,\be}\;;\;
(4)_\be\,\equiv\,h_{\ep\al ,\be}\B h^{\ep\al}\;;\nn\\
\, & \,(5)_\be\,\equiv\,h_{\ga\de ,\be\al}\B h^{\ga\de ,\al}\;;\;
(6)_\be \,\equiv\,h_{\al\be ,\ga\de}\,h^{\ga\de ,\al}\;;\;
(7)_\be\,\equiv\, h_{\al\de ,\be\ga}\,h^{\ga\de ,\al}\;;\;
(8)_\be\,\equiv\, h_{\be\de ,\al\ga}\,h^{\ga\de ,\al}\;;\nn\\
\, & \,(9)_\be \,\equiv\,h_{\ep\be}\B b^\ep\;;\;
(10)_\be\,\equiv\,b^\ep\B h_{\be\ep}\;;\;
(11)_\be\,\equiv\,b_{\be,\ga\de}\,h^{\ga\de}\;;\;
(12)_\be\,\equiv\,b_{\de ,\ga\be}\,h^{\ga\de}\;;\nn\\
\, & \,(13)_\be \,\equiv\,{b_\ga}^{,\ga\al}\,h_{\al\be}\;;\;
(14)_\be\,\equiv\,b^{\ep ,\sigma}\,h_{\be\ep ,\sigma}\;;\;
(15)_\be\,\equiv\,b^{\ga ,\s}\,h_{\be\s ,\ga}\;;\;
(16)_\be\,\equiv\,b^{\de ,\ga}\,h_{\ga\de ,\be}\;;\nn\\
\, & \,(17)_\be \,\equiv\,\B c^\al h_{\al\be}\;;\;
(18)_\be \,\equiv\,c^\al \B h_{\al\be}\;;\;
(19)_\be \,\equiv\,c^{\ep ,\al}\,h_{\ep\be ,\al}\;;\;
(20)_\be\,\equiv\,c^{\s ,\al}\,h_{\al\s ,\be}\;;\nn\\
\, & \,(21)_\be \,\equiv\, c_{\ep ,\al\be}\,h^{\ep\al}\;;\;
(22)_\be\,\equiv\,b_\be\,{b_\ga}^{,\ga}\;;\;
(23)_\be\,\equiv\, b_\ga\,{b^\ga}_{,\be}\;;\;
(24)_\be\,\equiv\,b_\ga\,{b_\be}^{,\ga}\;;\nn\\
\, & \,(25)_\be\,\equiv\,c^\al\,b_{\be ,\al}\;;\;
(26)_\be\,\equiv\,c^\al\,b_{\al ,\be}\;;\;
(27)_\be\,\equiv\,c_{\ep ,\be}\,b^\ep\;;\;
(28)_\be\,\equiv\,{c_\de}^{,\de}\,b_\be\;;\nn\\
\, & \,(29)_\be\,\equiv\,c_\be\,{b_\al}^{,\al}\;;\;
(30)_\be\,\equiv\,c^\al\,c_{\al ,\be}\;;\;
(31)_\be\,\equiv\,{c_\al}^{,\al}\,c_\be \;;\;
(32)_\be\,\equiv\,h\,\B b_\be\;;\nn\\
\, & \,(33)_\be\,\equiv\,h\,{b^\al}_{,\al\be}\;;\;
(34)_\be\,\equiv\,h \,\B c_\be\;, \qquad\nn
\ea

\nind    it can de demonstrated that

\ba 
-[\mu\rho ,\be ]\,\, {{\delta L_{F.P}}\over{\delta h^{\rho\mu}}}=\, \lambda
\Bigl[\,2( 1)_\be-\,(4 )_\be-2\,(14 )_\be-2\,(15 )_\be+2\,(16)_\be+ &2\,( 19)_\be-\,(20 )_\be
+\nn\\ 
+2\,( 22)_\be 
-2\,( 28)_\be-\,(29 )_\be+\,(31 )_\be \,\Bigr]\;,&\nn\\
\ea

\vs

In a similar way, some lengthy algebra leads to

\ba
\eta_{\mu\be}\,\,\Biggl({{\delta L_{3}}\over{\delta h^{\rho\mu}}}\Biggr)_{,\rho}=\,
\Biggl[-x_1-x_4+{1\over 2} x_5\Biggr]\,(1)_\be+\Biggl[-x_1-{1\over 2}x_5
\Biggr] \,(2)_\be +\nn\\
+\Biggl[-x_2-x_{10}\Biggr]\,(3)_\be+\Biggl[x_6-x_2 - x_{10} \Biggr]\,(4)_\be+
\Biggl[x_6-2x_2+2 x_7 -2x_{10}\Biggr]\,(5)_\be+\nn\\
+\Biggl[-2x_6-x_5+ x_4\Biggr]\,(6)_\be+
\Biggl[-x_4-2x_9\Biggr]\,(7)_\be+
\Biggl[-x_4-x_1\Biggr]\,(8)_\be+\nn\\
+\Biggl[-x_1-x_4-x_3\Biggr]\,(9)_\be+
\Biggl[-x_1-x_3-{1\over 2} x_5\Biggr]\,(10)_\be+
\Biggl[-2x_6-{1\over 2} x_5\Biggr]\,(11)_\be+\nn\\
+\Biggl[-x_4-x_3-{1\over 2} x_5 -x_9 -x_{11}\Biggr]\,(12)_\be+
\Biggl[-x_2-{1\over 2} x_5\Biggr]\,(13)_\be+
\Biggl[-x_1-2 x_3-x_4-{1\over 2} x_5\Biggr]\,(14)_\be+\nn\\
+\Biggl[x_4-{3\over 2} x_5 -2x_6\Biggr]\,(15)_\be+
\Biggl[-x_3-x_4 - x_9 -x_{11}\Biggr]\,(16)_\be+
\Biggl[-{1\over 2}x_9-x_{10}-{1\over 2} x_{11}\Biggr]\,(17)_\be+\nn\\
+\Biggl[-{1\over 2}x_{11}+{1\over 2} x_9-2x_7\Biggr]\,(18)_\be+
\Biggl[-2x_7-x_{11}\Biggr]\,(19)_\be+
\Biggl[-2x_{13}-{1\over 2}x_{11}+{1\over 2} x_9\Biggr]\,(20)_\be+\nn\\
+\Biggl[-{1\over 2}x_9-{1\over 2} x_{11}-2x_{13}\Biggr]\,(21)_\be+
\Biggl[-x_2+x_3\Biggr]\,(22)_\be+
\Biggl[-x_4-2x_3-{1\over 2} x_5+2x_8 -2x_{11}\Biggr]\,(23)_\be+\nn\\
+\Biggl[-2x_6-{1\over 2} x_5+x_3\Biggr]\,(24)_\be+
\Biggl[-2x_7-2x_8+{1\over 2} x_{11}-{1\over 2}x_9\Biggr]\,(25)_\be+\nn\\
+\Biggl[-{1\over 2}x_{11}+{1\over 2} x_{9}-x_8-2x_{13}\Biggr]\,(26)_\be+
\Biggl[-x_9-x_8-2 x_{13}\Biggr]\,(27)_\be +\nn\\
+\Biggl[-x_8-{1\over 2}x_9+{1\over 2} x_{11}-x_{10}\Biggr]\,(28)_\be+
\Biggl[-x_8-{1\over 2}x_9+{1\over 2} x_{11}-x_{12}\Biggr]\,(29)_\be+\nn\\
+\Biggl[x_{13}-2x_{12}-2 x_{14}\Biggr]\,(30)_\be+
\Biggl[x_{13}-x_{12}-2 x_{14}\Biggr]\,(31)_\be+
\Biggl[-x_8-2x_7 \Biggr]\,(32)_\be+\nn\\
+\Biggl[-x_8-x_{12}\Biggr]\,(33)_\be+
\Biggl[-x_{12}-2x_{14}\Biggr]\,(34)_\be\;.\nn\\
\ea

\vs

   Equating terms in (A.1) and (A.2) we obtain a system of 34 equations and 14 unknowns:

\ba
-x_1-x_4+{1\over 2} x_5\,=\,2\lambda\nn\\
-x_1-{1\over 2}x_5\,=\,0\nn\\
-x_2-x_{10}\,=\,0\nn\\
...\qquad =...\nn\\
-x_8-x_{12}\,=\,0\nn\\
-x_{12}-2x_{14}\,=\,0\;.\nn\\
\ea

\vs

Despite of the much bigger number of equation than unknowns, system (A.3) is 
compatible, being the coefficients appearing in (4.12) it´s solution:

\ba x_1\,=\,-2\,\lambda\,,\quad
x_2\,=\,-2\,\lambda\,,\quad
x_3\,=\,0\,\quad
x_4\,=\,2\,\lambda\,,\nn\\
x_5\,=\,4\,\lambda\,,\quad
x_6\,=\,-\,\lambda\,,\quad
x_7\,=\,{1 \over 2}\,\lambda\,,\quad
x_8\,=\,-\,\lambda\,,\nn\\
x_9\,=\,-\,\lambda\,,\quad
x_{10}\,=\,2\,\lambda\,,\quad
x_{11}\,=\,-3\,\lambda\,,\quad
x_{12}\,=\,\,\lambda\,,\nn\\
x_{13}\,=\,\,\lambda\,,\quad
x_{14}\,=\,-{1 \over 2}\,\lambda\;.\nn
\ea

\ms

 {\large {\bf{B.}}  {\ul{Some calculations concerning Condition B.}}} 
\setcounter{section}{2}
\setcounter{equation}{0}

\ms

We now set out the resolubility of (5.6); beginning with it's lowest order version (5.7). 
The first step is to determine $\Bigl(T^{(2)}_{\al\be}\Bigr)^{,\,\al}$ using the machinery 
developed in section 2.\vs

In this direction we write the infinitesimal Poincare variation of the field

\ba \delta h^{\prime\al\be}\,(x^\prime )\,=\,{1\over 2}\,\omega_{\mu\nu}
\,\bigl( S^{\mu\nu} {\bigr)}^{\,\al\be}_{\,\ep\tau}\,\, h^{\ep\tau}(x)\;,\nn\ea

\nind where

\ba \bigl( S^{\mu\nu} {\bigr)}^{\,\al\be}_{\,\ep\tau}\,=\,
{1\over 2}\,\Biggl( \,\delta^\al_\ep\,\eta^{\mu\be}\,\delta^\nu_\tau -
\delta^\al_\ep\,\eta^{\nu\be}\,\delta^\mu_\tau
+\delta^\be_\ep\,\eta^{\mu\al}\,\delta^\nu_\tau
-\delta^\be_\ep\,\eta^{\nu\al}\,\delta^\mu_\tau +\nn\\
+\eta^{\mu\al}\,\de^\nu_\ep\,\de^\be_\tau
-\eta^{\nu\al}\,\de^\mu_\ep\,\de^\be_\tau
+\eta^{\mu\be}\,\de^\nu_\ep\,\de^\al_\tau 
-\eta^{\nu\be}\,\de^\mu_\ep\,\de^\al_\tau \, \Biggr)\;, \nn
\ea

\nind and according (2.3)

\ba f^{\mu\rho\nu}\,=\,{1\over 2}\,\Biggl( L^\mu_{\al\be}\,
\bigl( S^{\rho\nu} {\bigr)}^{\,\al\be}_{\,\ep\tau}\,
+\,L^\rho_{\al\be}\,
\bigl( S^{\nu\mu} {\bigr)}^{\,\al\be}_{\,\ep\tau}\,
-\,L^\nu_{\al\be}\,
\bigl( S^{\mu\rho} {\bigr)}^{\,\al\be}_{\,\ep\tau}\,
\Biggr)\, h^{\ep\tau}\nn\ea

\ba L^\mu_{\al\be}\,\equiv\,
{\partial L_F\over{\partial h^{\al\be}_{,\,\mu}}}\;. \nn\ea
\vs

Substituting these values in (2.2) it is not hard to see that

\ba -\lambda\,\Bigl(T^{(2)}_{\al\be}\Bigr)^{,\,\al}\,=\,
-\lambda\,\Biggl[ -2\,(1)_\be+2\,(2)_\be +2\,(4)_\be+2\,(10)_\be
-2\,(11)_\be-2\,(12)_\be+2\,(13)_\be +\nn\\
+2\,(14)_\be+2\,(15)_\be-4\,(16)_\be-2\,(17)_\be-2\,(19)_\be+2\,(20)_\be +\nn\\
+2\,(21)_\be-2\,(23)_\be-2\,(24)_\be+2\,(27)_\be+2\,(29)_\be-2\,(31)_\be \;\Biggr]\;.\nn\\
\nn\\
\ea

\vs 

  Equating now terms in (A.7) and (B.1) we obtain the new system of equations

\ba
-x_1-x_4+{1\over 2} x_5\,=\,2\lambda\nn\\
-x_1-{1\over 2}x_5\,=\,-2\lambda\nn\\
-x_2-x_{10}\,=\,0\nn\\
...\qquad =...\nn\\
-x_8-x_{12}\,=\,0\nn\\
-x_{12}-2x_{14}\,=\,0\;.\nn
\ea
\nind Which is incompatible. Anyway we have not considered yet the the implications of $h_{\al\be}
\in H^*$.\vs

We introduce the notation

\ba 
-[\mu\rho ,\be ]\, {{\delta L_{F.P}}\over{\delta h^{\rho\mu}}}
\equiv\,\sum_{i=1}^{34}\,\Omega_i\,(i)_\be\,+\vartheta\,(\geq 3)\nn\ea

\ba \eta_{\mu\be}\,\Biggl({{\delta L_{3}}\over{\delta h^{\rho\mu}}}\Biggr)_{,\rho}
\equiv\,\sum_{i=1}^{34}\,z_i\,(i)_\be\,+\vartheta\,(\geq 3)\quad,\nn\ea

\nind (where $\vartheta (p)$ stands for the terms of order $p$ in the $h's$) and 
reexpress the second order version of the (5.5) cancelation as

\ba \sum_{i=1}^{34}\,z_i\,(i)_\be =\,\sum_{i=1}^{34}\,\Omega_i\,(i)_\be\;.
\ea

\vs
   Calling  

\ba -\lambda\,\Bigl(T^{(2)}_{\al\be}\Bigr)^{,\,\al}\,
\equiv\,\sum_{i=i}^{34}\,\Theta_i\,(i)_\be\nn
\ea

\nind we rewrite the incompatible condition (5.4) as

\ba \sum_{i=1}^{34}\,z_i\,(i)_\be\,=\,\sum_{i=1}^{34}\,\Theta_i\,(i)_\be\quad.\ea

\vs

Dealing with the resolubility of Condition B we still have to impose the condition 
$h_{\al\be}\in H^*$, {\it{i.e.}}
the 34 $(i)_\be$ are not longer independent. We can use (B.2) to write $(j)_\be\,;\; 1\leq j \leq 34$
as a linear combination of the 33 remainig terms, substitute in (B.3) and obtain a new
system of equations. In this direction, we say that a set of values $\{\bar{x}_i\}_{i=1}^{14}$
is a solution to our problem if, finding the expression (B.2) for $\{{x}_i\}_{i=1}^{14}\,=\,\{\bar{x}_i\}_{i=1}^{14}$ and
replacing  any of the $(k)_\be$ there appearing in (B.3), it identically 
vanishes. In other words, we have to deal with $34$ different systems of equations, depending on the $(j)_\be$
replaced. We shall distinguish two cases\vs

 i) If the replaced $(j)_\be$ is such that $\Theta_j\,\not= ,\Omega_j$, when identifying
the coefficients of $(i)_\be ,,\; i \not= j$; a new system 
(33 equations, 14 unknowns) is obtained. \vs

It is not hard to check that the new system of equations is given by

\ba z_i\,\bigl(x_1,x_2,...,x_{14}\bigr)\,=\,{{\Omega_i-\Theta_i}\over{\Omega_j-\Theta_j}}
\,z_j\,\bigl(x_1,x_2,...,x_{14}\bigr)+
{{\Omega_j\Theta_i-\Theta_j\Omega_i}\over{\Omega_j-\Theta_j}}\nn\\
j \;fixed \; ;\qquad \qquad i \not= j\;,\nn
\ea

\nind and, independently of the $(j)_\be$ replaced, the resolution of the system gives 
a solution which can be parametrized as

\ba x_1\,=\,-2\,\lambda\,+\,\ga\,,\quad
x_2\,=\,-2\,\lambda\,+\,\ga\,,\quad
x_3\,=\,0\,,\quad
x_4\,=\,2\,\lambda\,-\,\ga\,,\nn\\
x_5\,=\,4\,\lambda\,,\quad
x_6\,=\,-\,\lambda\,-\,{1\over 2}\ga\,,\quad
x_7\,=\,{1 \over 2}\,\lambda\,+\,{1\over 4}\,\ga,\quad
x_8\,=\,-\,\lambda\,-\,{1\over 2}\,\ga\,,\nn\\
x_9\,=\,-\,\lambda\,+\,{1\over 2}\,\ga\,,\quad
x_{10}\,=\,2\,\lambda\,-\,\ga\,,\quad
x_{11}\,=\,-3\,\lambda\,-{1\over 2}\,\ga,\quad
x_{12}\,=\,\,\lambda\,+\,{1\over 2}\,\ga,\nn\\
x_{13}\,=\,\,\lambda\,+\,{1\over 2}\,\ga,\quad
x_{14}\,=\,-{1 \over 2}\,\lambda\,-\,{1\over 4}\,\ga.\nn\\ 
\ea  
\vs

Redefining the parameter $\ga$ as $\ga\equiv\,\ep\,\lambda$ we identify the coefficients
appearing in (5.8). It should be remarked that if we substitute the values (B.4) in (B.2), all
the $(j)_\be$ such that $\Theta_j\,=\,\Omega_j$ decouple from the expression. So,we conclude
that (B.4) is a solution to our problem.\vs

ii) If  $(j)_\be$ is such that $\Theta_j\,=\,\Omega_j$, when replacing it
all the contributions in $x_1,...,x_{14}$ cancel and what is obtained is nothing
but a linear condition on the $(i)_\be\,$ which is precisely (B.2) using (B.4). We conclude that
(B.4) is not only a solution, but the only solution to our problem.\vs

{\large {\bf
 Bibligraphy.}}

\ms
[1] Kraichnan.Phys. Rev. {\bf 98} (1955) 1118.
\vs

[2] Gupta S. Phys. Rev. {\bf 96} (1954) 1683,

    \hs Gupta S. Rev. Mod. Phys. {\bf 29} (1957) 334.
\vs

[3] Feynman R. Lectures on gravitation. Penguin books.
\vs

[4] Deser, Gen. Rel. and Grav. {\bf 1} (1970) 9.
\vs

[5] Barut A.O. Electrodynamics and classical theory of fields and particles. Dover
Publications, INC. New York.
\vs

[6] Weinberg S. Phys. Rev. {\bf 138} (1965) 988.
\vs

[7] L\'opez-Pinto A. General Covariant space-time theories.
In preparation.
\vs

[8] See for instance:\\

\nind - Thirring W.E., Fortsch. Phys. {\bf 7} (1959) 79.\\
- Thirring W.E. Ann. Phys. (N.Y.) {\bf16} (1961) 96.\\
- Landau L.D. and Lifshitz E.M. The classical theory of fields, Moscow 1973, etc.\\
-Weinberg S. Gravitation and Cosmology. New York, 1972.

\end{document}